\documentclass[prl, twocolumn,superscriptaddress]{revtex4}
\usepackage{amsmath,amssymb,bm}
\usepackage{graphicx}
\usepackage{epstopdf}
\usepackage{latexsym}
\usepackage{subfigure}
\usepackage{color}
\usepackage{natbib}
\usepackage{braket}
\usepackage{hyperref}
      
\begin{document}

\title{SO($N$) singlet-projection model on the pyrochlore lattice}

\author{Matthew S. Block}
\affiliation{Department of Physics \& Astronomy, California State University, Sacramento, CA 95819}

\author{Jared Sutton}
\affiliation{Revature, Reston, VA 20190}

\begin{abstract}
We present an extensive quantum Monte Carlo study of a nearest-neighbor, singlet-projection model on the pyrochlore lattice that exhibits SO($N$) symmetry and is sign-problem-free. We find that in contrast to the previously studied two-dimensional variations of this model that harbor critical points between their ground state phases, the non-bipartite pyrochlore lattice in three spatial dimensions appears to exhibit a first-order transition between a magnetically-ordered phase and some, as yet uncharacterized, paramagnetic phase.  We also observe that the magnetically-ordered phase survives to a relatively large value of $N=8$, and that it is gone for $N=9$.
\end{abstract}
\date{\today}
\maketitle

\section{Introduction}
The search for exotic quantum phase transitions in two dimensions has been a fruitful endeavor from the perspective of numerical investigations. Of note is the prediction, detection, and characterization of so-called deconfined quantum critical points (DQCPs). These critical points defy traditional Landau-Ginzburg-Wilson theory of phase transitions by allowing for a direct, continuous transition between two phases that break fundamentally different symmetries. Furthermore, numerical evidence corroborated the claim that at the critical point, an emergent U(1) gauge field mediates interactions between spinon degrees of freedom that are normally confined in the adjacent phases.\cite{senthil2004:science} The numerical linchpin to the success of these studies was the development of SU($N$)-symmetric spin-singlet-projection models deployed on several different bipartite two-dimensional (2D) lattices,\cite{sandvik2007:deconf,lou2009:sun,kaul2012:j1j2} which allowed for comparison to the large-$N$ calculations on the descriptive gauge field theory, a non-compact CP$^{N-1}$ field theory.~\cite{motrunich2004:hhog,senthil:dccd} These SU($N$) models also attracted some attention from the experimental community for their relevance to optical lattices of certain fermionic alkaline-earth atoms. The nuclear spins in these systems exhibited SU($N$) symmetry with $N$ as large as 10.~\cite{gorshkov2010:sun}

A natural extension was to consider the same type of sign-problem-free operator on a non-bipartite lattice, such as the triangular~\cite{kaulTriangular} or kagome,~\cite{kagome} where the symmetry is SO($N$). Both of these studies showed evidence of exotic, DQCP-like critical points separating the phases as well as the presence of spin-liquid phases. While the identification and characterization of these condensed matter phenomena are interesting in their own right, Demler \emph{et. al.} describe a SO(5) theory that connects antiferromagnetism and superconductivity, phases often seen adjacent in high-temperature superconducting cuprates, heavy fermion compounds, and organic superconductors, thus lending some experimental relevance to the work presented herein.\cite{demler:so5}

An obvious question is whether these critical transitions persist into three dimensions where previous studies of a similar nature have seen them lost to comparatively plain first-order transitions.\cite{cubic} Here we take the first step in these investigations by deploying the same SO($N$)-symmetric nearest-neighbor singlet-projection model, augmented by a next-nearest-neighbor permutation term, on the three-dimensional (3D), non-bipartite pyrochlore lattice.

\section{Model}
We consider the pyrochlore lattice where each site has a Hilbert space of $N$ states, denoted for site $j$ as $\Ket{\sigma}_j$ where $\sigma=1,\ldots,N$. We will refer to this as the \emph{color} of the spin. By using the fundamental representation of SO($N$) on each site it is possible to construct spin singlets on any two sites: $\Ket{S_{ij}}=\frac{1}{\sqrt{N}}\sum_\sigma\Ket{\sigma\sigma}_{ij}$. We can then construct the singlet-projection operator for a pair of sites as $\hat{\mathcal{P}}_{ij}=\Ket{S_{ij}}\Bra{S_{ij}}$.  This follows closely the previous numerical studies on the triangular lattice~\cite{kaulTriangular} and kagome lattice~\cite{kagome}. We consider this operator acting on nearest neighbors of the lattice and this is the first term in our model Hamiltonian:
\begin{equation}
\label{eqn:J1term}
\hat{\mathcal{H}}_{J_1}=-J_1\sum_{\braket{ij}}\hat{\mathcal{P}}_{ij}
\end{equation}
with $J_1>0$ in all cases. We can study this model on its own for integer values of $N$ to map out the phase diagram as a function of the symmetry order (see Results below). To gain a more detailed understanding of the phase transition between observed phases, we can add a second term that acts on the shortest bonds joining sites on the same sublattice (denoted $\{ij\}$; see Fig.~\ref{fig:pyrochlore}):
\begin{figure}[t]
\centerline{\includegraphics[width=\columnwidth]{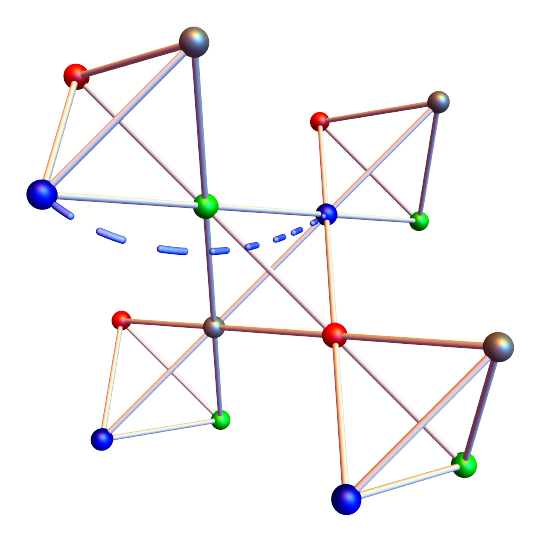}}
\caption{(color online).  A small cluster of the pyrochlore lattice ($L=4$). Sites on the each of the four sublattices are shown in the same color. The solid lines connect nearest-neighbor sites where the coupling is $J_1$. A single representative dotted line is shown connecting nearest-neighbor sites \emph{on the same sublattice} where the coupling is $J_2$. All such next-nearest-neighbor sites are coupled in the same way.}
\label{fig:pyrochlore}
\end{figure}

\begin{equation}
\label{eqn:J2term}
\hat{\mathcal{H}}_{J_2}=-J_2\sum_{\{ij\}}\hat{\Pi}_{ij},
\end{equation}
where $\hat{\Pi}_{ij}=(1/N)\sum_{\sigma,\eta}\Ket{\sigma\eta}_{ij}\Bra{\eta\sigma}_{ij}$, the so-called permutation operator, which encourages magnetic ordering for $J_2>0$.  Our full model is thus $\hat{\mathcal{H}}=\hat{\mathcal{H}}_{J_1}+\hat{\mathcal{H}}_{J_2}$.  In the $J_2$-only model for any finite $N$, each sublattice would be perfectly, but independently, magnetically ordered. By turning on a small $J_1$ at that point, the sublattices couple together and all spins align.  Therefore, if we start in the paramagnetic phase for some large $N$, there must exist some $g\equiv J_2/J_1$ beyond which magnetic order is restored.  By varying $g$, we can continuously tune from one phase to the other and perform a detailed study of the properties of the phase transition.

\section{Method and Measurements}
In all cases, we study lattices with $N_\text{spin}=L^3/4$ sites where $L$ is the side length of the cubic lattice in which we could inscribe our section of pyrochlore. Periodic boundary conditions are enforced along all three standard axes, which preserves the rotational symmetry of the lattice. We employ the stochastic series expansion (SSE) method for our quantum Monte Carlo (QMC), which samples via local bond updates and global loop updates. Aside from some proprietary measurement code and the generalization to an arbitrary symmetry order $N$, the QMC algorithm was developed and described in detail by Anders Sandvik.~\cite{sandvik2010:vietri}

Here we will summarize the key features of the SSE method. The quantum Boltzmann factor that shows up in the partition function is first Taylor expanded:
\begin{equation}
\label{eqn:sse}
e^{-\beta\hat{\mathcal{H}}}=\sum_{n=0}^\infty \frac{(-\beta\hat{\mathcal{H}})^n}{n!},
\end{equation}
where $\beta$ is the usual reciprocal temperature. Next, we insert the model Hamiltonian defined above as the sum of Eqs.~(\ref{eqn:J1term}) and (\ref{eqn:J2term}). We can then imagine applying the exponent $n$ to the sums of terms in those equations generating products with varying numbers of bond operators of the form $\Ket{\sigma\eta}_{ij}\Bra{\eta\sigma}_{ij}$, which we can refer to as \emph{operator strings}.  The strings themselves can be visualized as a $(3+1)$-dimensional spacetime where the ``time" dimension is so-called imaginary time, $\tau$, and flows from one operator in the string to the next. The overwhelming majority of these strings will yield zero due to orthogonality of the spin states and many more will be zeroed out by the application of the trace in the partition function:
\begin{equation}
Z=\mathrm{Tr}\,e^{-\beta\hat{\mathcal{H}}}=\sum_\alpha \Braket{\alpha|e^{-\beta\hat{\mathcal{H}}}|\alpha},
\end{equation}
where $\Ket{\alpha}$ can, in principle, be any orthonormal basis. In our case, we use a simple direct product of the spin states on each site denoted by the color of the spin on that site: $\Ket{\sigma_1\sigma_2\ldots\sigma_{N_\text{spin}}}$, where $\sigma_i$ can vary from 1 to $N$.

The job of an effective SSE QMC algorithm is to sample, not only over all possible states $\Ket{\alpha}$, but, critically, over only operator strings that will yield non-zero values in the partition function while preserving ergodicity. To this end, we start with an empty spacetime --- no operators in the string --- and some random state $\Ket{\alpha}$, which must be realized at both $\tau=0$ and $\tau=\beta$. We implement two kinds of sampling updates. First, a diagonal update, which allows for the insertion or removal of operators as mentioned in the preceding paragraph with $\sigma=\eta$. This kind of update leaves the state $\Ket{\alpha}$ unchanged, but increases or decreases the length of the operator string. This corresponds to sampling over the exponent $n$ in Eq.~(\ref{eqn:sse}). There is a weight associated with a given spacetime configuration with $n$ operators in the string and these weights are used with a standard Metropolis algorithm to accept or reject the diagonal updates.  The weights are manifestly built from the inner products between two-site states as they show up in the model Hamiltonian. The explicit minus signs in Eqs.~(\ref{eqn:J1term}) and (\ref{eqn:J2term}) with both $J_1$ and $J_2$ strictly non-negative, along with the minus sign preceding $\beta$ in the Boltzmann factor, ensure that all weights for configurations generated via diagonal updates will be strictly positive, thus averting the dreaded sign problem.

Before describing the second type of sampling update, we should characterize the type of magnetic ordering that is observed with this model.  Classically, magnetic order would look like all spins having the same color, a variant of ferromagnetism.  To measure this in the quantum system, we \emph{could} introduce a magnetic order parameter of the form
\begin{equation}
\hat{Q}_{\sigma\eta}=\Ket{\sigma}\Bra{\eta}-\frac{\delta_{\sigma\eta}}{N},
\end{equation}
which, due to its tensorial nature, invites the characterization of the magnetic order as ``quadrupolar." While this is conceptually useful, we will use a different magnetic order parameter, as described below, to perform our measurements numerically.

The diagonal update by itself cannot achieve ergodicity since it cannot change the state $\Ket{\alpha}$. This is accomplished via global loop updates. As the spacetime fills with operators, closed loops connecting spins of the same color will form if we allow the loops to wrap around temporally, meaning they pass through the state at $\tau=0$. Thus, if we change the color of all spins on a given loop to some random other color, it will update $\Ket{\alpha}$ and create off-diagonal operators, which also must be allowed in our operator strings. The loop updates leave the number of operators in the string, $n$, unchanged and hence the weights of the current and proposed configurations are always equal so the loop updates are always accepted.  Together, these two sampling updates can realize every spacetime configuration with non-zero weight thus achieving ergodicity, at least in principle (see below for a discussion of the limits of the sampling algorithm when $N$ is large). If this is hard to visualize, we invite the reader to consult Sandvik's original SSE work~\cite{sandvik2010:vietri}, which contains many helpful figures.

The loops built by the SSE algorithm live within a $(3+1)$-dimensional spacetime the size of which scales as $L^3\beta$. They connect spins of the same color and, as such, the proliferation of these loops throughout the lattice speaks to the degree of long-range magnetic ordering. More precisely, if the loops wrap around the entire lattice, we call this a \emph{winding} and the total number of times this occurs among all loops along one of the three standard axes constitutes the \emph{winding number} for that direction:
\begin{equation}
W_\mu = \sum_{i=1}^{\text{\# of loops}}\frac{\text{displacement of loop $i$ along $\mu$-axis}}{L}
\end{equation}
for $\mu=x,y,z$. This turns out to be an extensive quantity since the spacetime grows with lower temperature (larger $\beta$), but, in three dimensions, we can define the intensive \emph{spin stiffness} along each direction:
\begin{equation}
\rho_{s,\mu}\equiv\frac{\Braket{W_\mu^2}}{\beta\,L}
\end{equation}
and this is the order parameter we shall use to detect the presence of magnetic order numerically.  Given the symmetry of the lattice, we expect to have $\rho_{s,x}=\rho_{s,y}=\rho_{s,z}$ and can therefore look at any one component or average across all three components to improve the quality of our statistical estimates.

\section{Results}
\subsection{The $J_1$-only Model and the Critical Value of $N$}
\begin{figure}[t]
\centerline{\includegraphics[width=\columnwidth]{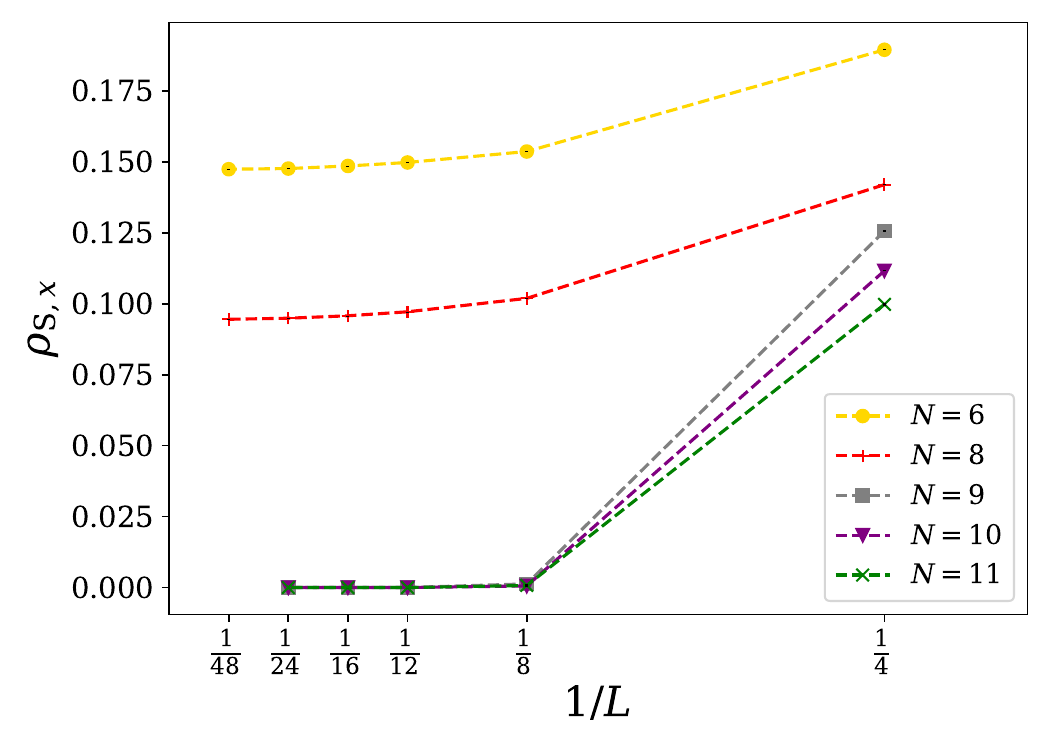}}
\caption{(color online).  The magnetic order parameter, $\rho_{\mathrm{s},x}$, as a function of reciprocal system size for various values of $N$ for the $J_1$-only model. For sufficiently large $L$, we conclude that magnetic order survives up through $N=8$ and is destroyed for larger $N$. We use $\beta=10L$ for this part of the study. Error bars are too small to be visible.}
\label{fig:rhoSvsN}
\end{figure}

We began by studying the $J_1$-only model while varying $N$ to determine how the symmetry order affected the realized phase.  A modest investigation using system sizes $L=4, 8, 12, 16, 24, 48$ (Fig.~\ref{fig:rhoSvsN}) revealed that magnetic order persists for $N\leq8$ and vanishes for $N\geq9$. We note that $N=9$ appeared to be just on the paramagnetic side of the transition, but very close to it, such that it suffered from the ergodicity issue we will describe below. A similar resolution to what is described therein involving a comparison of energies allowed us to conclude the $N=9$ was indeed paramagnetic.

\subsection{The Ergodicity Problem}
Our next goal is to characterize the nature of the transition between the two known phases. To accomplish this, we start with a large value of $N$ on the paramagnetic side of the critical value --- in this case, we used $N=14$ --- and turn on the $J_2$ term. We used $\beta=6.25L$ for the remainder of this study, relaxing a bit from the $\beta=10L$ used in the $J_1$-only model above, which was perhaps a bit more than necessary.

To locate the transition, we considered a range of values of $g$ on system sizes $L=8, 12, 16, 24, 48$. 
However, it became clear early in this investigation that there was a breakdown of ergodicity in the Monte Carlo sampling. As $g$ was decreased, the stiffness appeared to have a finite value, but then, occasionally, it would drop off to a significantly lower value, or even zero, and then return to the trend of finite stiffness values for still lower values of $g$. These jumps in the middle of otherwise smooth data are tell-tale signs of inadequate equilibration and indeed a careful examination of the data files, which were binned, showed the value of stiffness dropping as additional bins were collected. However, some processor cores returned bins that didn't drop at all. Despite running millions of equilibration Monte Carlo sweeps, we were forced to conclude that a report of zero stiffness was trustworthy --- as it exhibited lower energy and stiffness only ever \emph{decreased} with additional bin collection --- while a report of finite stiffness could not be believed.

Already, this behavior is compelling evidence of the presence of a first-order transition. There appear to be two phases, close in energy, but quite distinct in character with a magnetic order parameter that exhibits a discontinuity in the neighborhood of the transition. But finding the location of the transition precisely is very challenging if we cannot pinpoint where the order parameter genuinely drops to zero due to the breakdown of ergodicity. It is a nightmare scenario for a QMC practitioner because we are already performing millions of equilibration sweeps at great computational cost for the larger systems sizes and there is no way of knowing how many more millions would be necessary to find the true ground state, or if that were even possible.

This breakdown of ergodicity is ultimately a consequence of the inadequacies of our sampling algorithm, which relies on both local spin updates and global loop updates within our spacetime. The loop updates are necessary for ergodicity, but are maximally efficient when the loops are long such that they update many spins at once. The function of the loop updates is therefore analogous to the famous Swendsen-Wang~\cite{swendsenwang} and Wolff~\cite{wolff:clusters} cluster updates for the Ising model (incidentally, our algorithm using the Swendsen-Wang version in choosing to update every loop in the spacetime with a random spin color each Monte Carlo sweep). The magnetically-ordered phase, with its long-range order, has a small number of very long loops joining spins of the same color, while the paramagnetic phase typically has a huge number of tiny loops each with a random color. In the paramagnetic phase, updating these loops is at once computationally costly and inefficient at updating the states. The development of a new process to augment our sampling algorithm that would more efficiently sample the paramagnetic phases has thus far eluded us.

Ergodicity issues like this can often be addressed by merely raising the temperature, but our analysis of energy versus temperature showed that we required rather small temperatures to capture the ground state behavior. One can try simply performing more equilibration sweeps, but, as mentioned above, we are already doing quite a lot and, even if we did significantly more, we could never be sure that we had indeed settled to the ground state near the transition point. Another popular technique is thermal annealing~\cite{kirkpatrick:thermalannealing} where the temperature starts high and is systematically lowered in an attempt to ``trap" the system in the global minimum of energy while avoiding any local minima of similar depth, but implementing this would have required a significant overhaul of our code and there existed a much quicker resolution that would also provide 100\% confidence in the outcome: an ad hoc form of variational Monte Carlo.  

An unintended consequence of the metastability of the two states within the QMC is that we are able to stay in one state or the other as we sweep past the transition point while varying $g$. On the magnetically-ordered (MO) side, our sampling algorithm will very efficiently find the lowest-energy state with finite stiffness by starting with a random spin state and empty spacetime (i.e., no loops initially); however, this may not be the true ground state as $g$ is lowered and the transition is passed.  Meanwhile, deep within the paramagnetic (PM) phase, our simulation, which, to be clear, is fully ergodic in principle, will find its way to a zero stiffness state (i.e., many tiny loops in the spacetime) and record the QMC spacetime configuration to a file. We can then copy this file for larger values of $g$ to use as a starting point. By lightly equilibrating to adjust for the different value of $g$, we can then find the lowest-energy state as we march to larger and larger values of $g$. But, again, ergodicity breaks down near the transition and the simulation will not easily find its way to the MO state until $g$ is raised significantly beyond the transition point.

In summary, starting with a random state will incorrectly tell us that magnetic order persists to a much lower value of $g$ than the transition value while starting with a PM state will tell us that the paramagnetism persists to a much higher value of $g$ than the transition value. But the true ground state must have the lowest energy; the process we used here is reminiscent of a variational Monte Carlo study~\cite{scherer:computationalphysics} in which an approximate ground state is determined by minimization of energy as parameters are varied within the proposed states. The famous shortcoming of this method is that one cannot be sure that the true ground state looks anything like those being proposed. Here, we do not have that problem; our ground state is either MO or PM, so whichever has the lower energy must be the true ground state.
\begin{figure}[t]
\centerline{\includegraphics[width=\columnwidth]{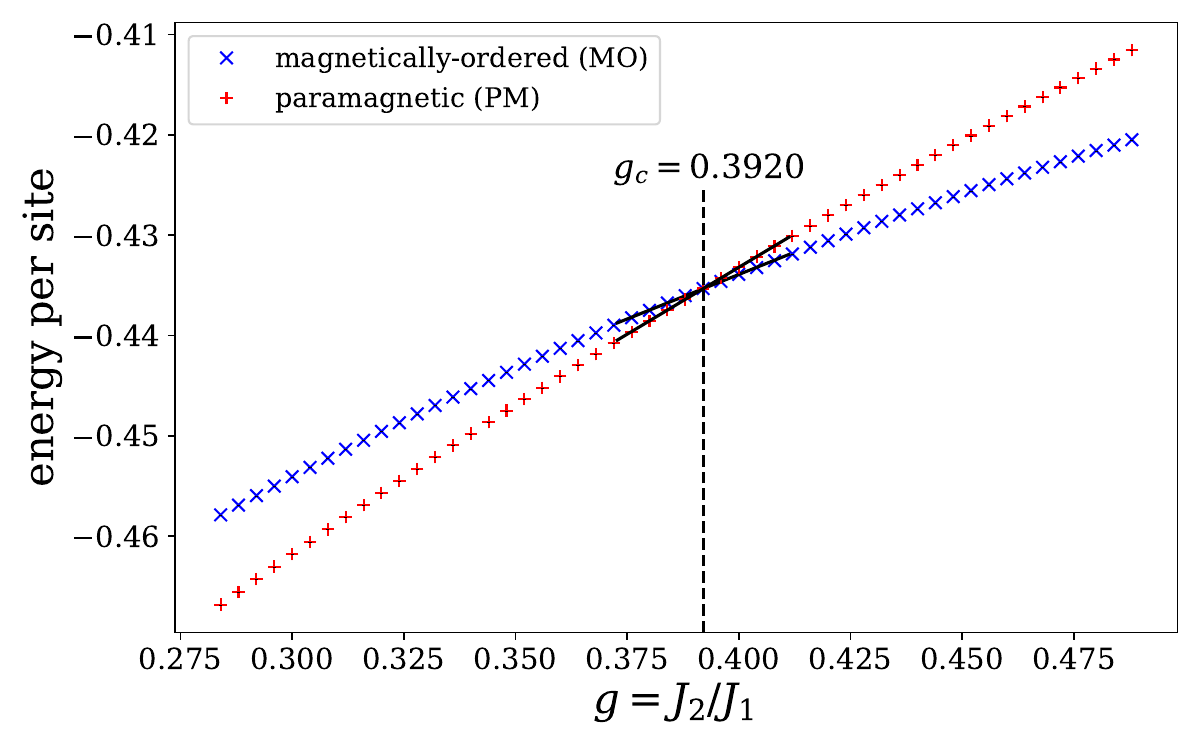}}
\caption{(color online).  The energy per site as a function of $g=J_2/J_1$ calculated from two different metastable states within the QMC for $L=24$ and $N=14$, the magnetically-ordered (MO) state and the paramagnetic (PM) state. The states differ qualitatively in the value of the magnetic order parameter --- finite for MO and zero for PM --- but yet they have very similar energies. The energy data in the vicinity of the crossover is fitted with two separate best-fit lines and the intersection point is calculated and reported as the transition value of $g$, which we call $g_c$. Error bars are too small to be visible.}
\label{fig:energyCross}
\end{figure}
\begin{figure}[t]
\centerline{\includegraphics[width=\columnwidth]{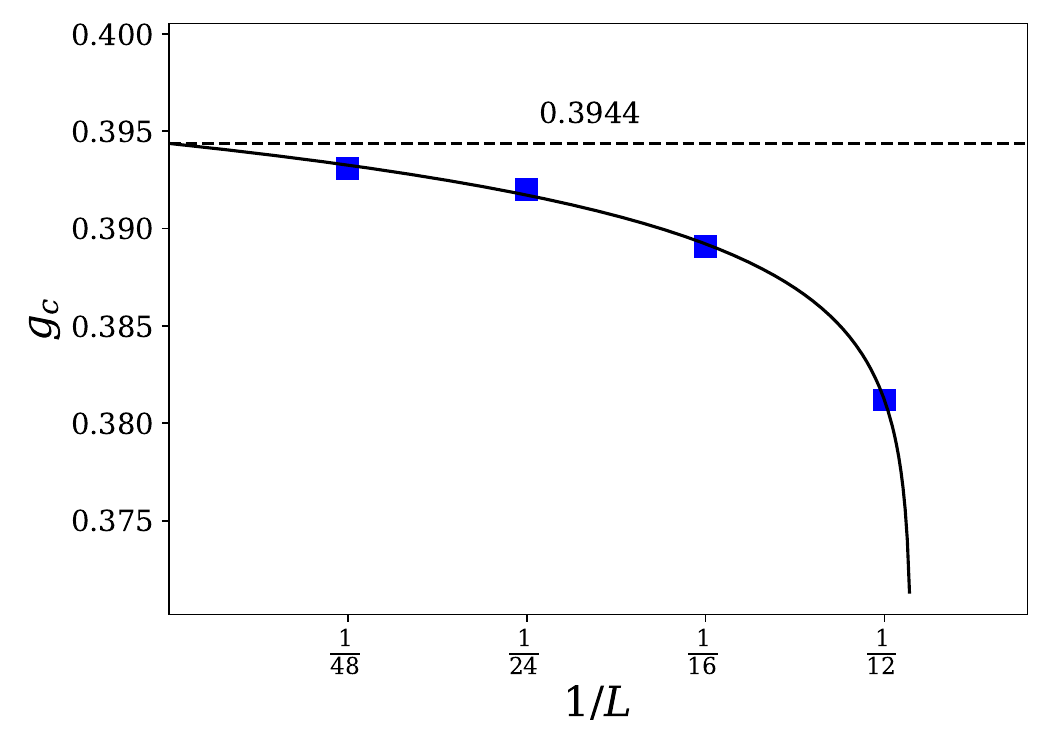}}
\caption{(color online).  The transition point drifts to larger values of $g$ as we increase system size suggesting that the magnetically-ordered phase is less stable for larger $L$ with fixed $N$. We fit a power law to the values of $g_c$ to extrapolate to the thermodynamic limit.}
\label{fig:gCvsL}
\end{figure}

With this strategy, we compare the energy of the MO and PM states as a function of $g$. The transition point, which we call $g_c$, occurs where the energies cross over; see Fig.~\ref{fig:energyCross} for an example of this for $L=24$. We repeated this energy crossover analysis for each of our system sizes and found the value of $g_c$ in each case. We plot these values versus $1/L$ in Fig.~\ref{fig:gCvsL} and determine that the value of $g$ in the thermodynamic limit is $g_{c,\infty}\approx 0.3944$ 
for $N=14$. Naturally, we expect this value to be less for smaller $N$ (still above $N=8$) and greater for larger $N$, though we did not investigate this in detail.
\begin{figure}[t]
\centerline{\includegraphics[width=\columnwidth]{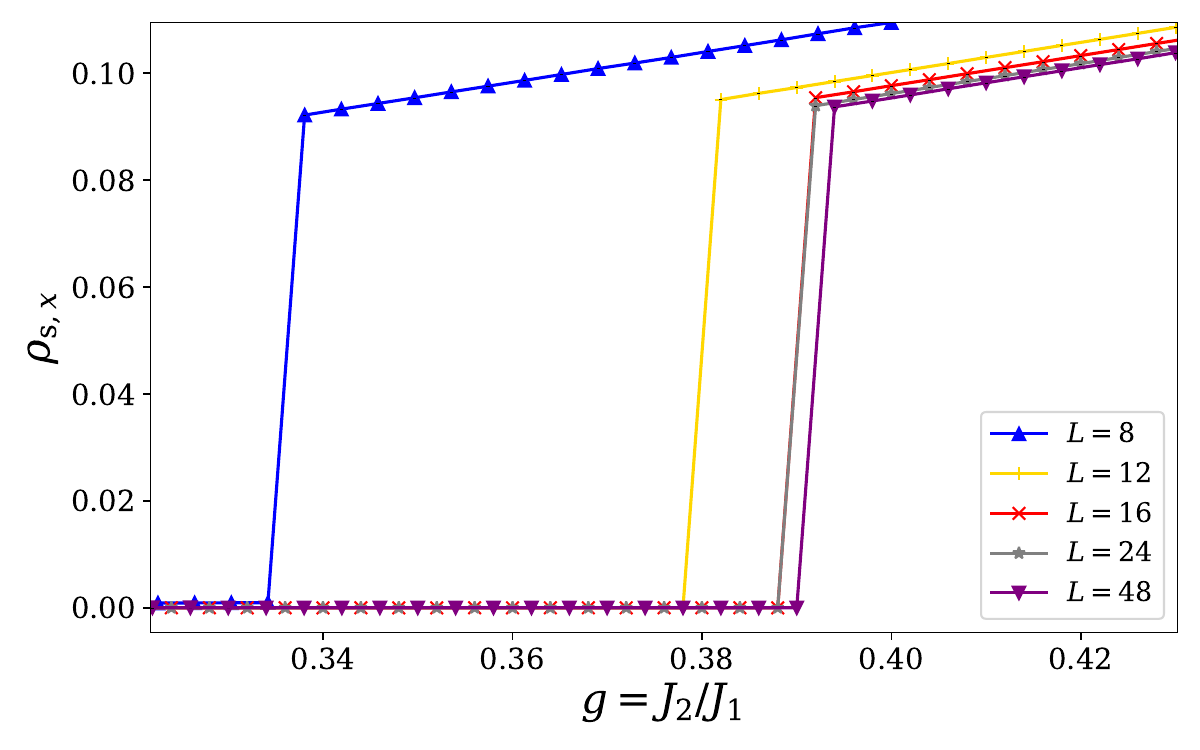}}
\caption{(color online).  The magnetic order parameter, $\rho_{\mathrm{s},x}$, as a function of $g$ for various system sizes. The sudden drop in the order parameter provides strong evidence of a first-order transition. Error bars are too small to be visible.}
\label{fig:rhoSvsG}
\end{figure}

\subsection{The Character of the Transition}
At last, we can return to the task of plotting the magnetic order parameter versus $g$ to visualize the transition on various system sizes; see Fig.~\ref{fig:rhoSvsG}. No longer are there artifacts of the inadequate equilibration due to the breakdown of ergodicity. The data points show a smooth trend on either side of the transition with invisible error bars and we can have certainty as to which data to use. Once we have determined the transition value of $g$ from the energy plots as described above, we can choose to use the PM stiffness data when $g<g_c$ and the MO data when $g>g_c$ for each system size. This is how the plot in Fig.~\ref{fig:rhoSvsG} was created. The sudden drop of stiffness from a finite value to zero across the transition, which persists on all system sizes studied, provides additional and compelling evidence of a first-order transition for this model on the pyrochlore lattice and that is our main result.

\section{Conclusion}
Our quantum Monte Carlo study of the SO($N$) singlet-projection model ($J_1$ only) on the pyrochlore lattice establishes the existence of quadrupolar magnetic order for $N\leq8$ and the destruction of this order for $N\geq9$. The augmented model with the continuously tunable parameter $J_2/J_1$ was studied in depth for $N=14$. While the stochastic series expansion algorithm is truly ergodic in all cases, its sampling efficiency wanes in the absence of magnetic order leading to an effective breakdown of ergodicity, especially near the transition point. Choosing the phase that minimizes the energy resolves this numerical challenge, allowing for a thorough investigation of the transition. Our study reveals the existence of a first-order transition separating the magnetically-ordered phase from the paramagnetic phase, as evidenced by a sharp, discontinuous drop in the magnetic order parameter. This result is in contrast to the wide range of studies of the same model on various two-dimensional lattices, which, with rare exception, all harbored exotic critical points, though it is consistent with other three-dimensional studies where critical behavior is lost and replaced by mundane first-order transitions.

There exists ample opportunity for extensions of these investigations.  For example, we avoided the critical value of $N$ where magnetic order first breaks down in the $J_1$-only model when studying the transition using the $J_1$-$J_2$ model, but one could certainly lower $N$ to see if any qualitative differences emerged, particularly for an odd value of $N$.  Of greater theoretical interest is to consider our same model on a non-trivial, bipartite, three-dimensional lattice where we return to SU($N$) symmetry, such as the diamond lattice.  The diamond lattice project is already underway and will serve as a logical capstone to this line of inquiry.

\section{Acknowledgements}
The authors wish to acknowledge and thank Ribhu Kaul (Penn State) who authored much of the original QMC code and suggested the pyrochlore lattice as an interesting application for the model studied herein. We gratefully acknowledge the National Science Foundation (NSF) and, in particular, the ACCESS collaboration (formerly XSEDE) and the San Diego Supercomputer Center whose Expanse cluster was instrumental to this study. The research reported here was supported in part by NSF DMR-130040. M. Block was additionally supported by Sacramento State's Department of Physics \& Astronomy Hu Research Award and Iloff Endowment Mentor Support Grant. J. Sutton was supported by Sacramento State's College of Natural Sciences \& Mathematics Summer Undergraduate Research Experience (SURE) Award.


\begin{thebibliography}{16}
\expandafter\ifx\csname natexlab\endcsname\relax\def\natexlab#1{#1}\fi
\expandafter\ifx\csname bibnamefont\endcsname\relax
  \def\bibnamefont#1{#1}\fi
\expandafter\ifx\csname bibfnamefont\endcsname\relax
  \def\bibfnamefont#1{#1}\fi
\expandafter\ifx\csname citenamefont\endcsname\relax
  \def\citenamefont#1{#1}\fi
\expandafter\ifx\csname url\endcsname\relax
  \def\url#1{\texttt{#1}}\fi
\expandafter\ifx\csname urlprefix\endcsname\relax\def\urlprefix{URL }\fi
\providecommand{\bibinfo}[2]{#2}
\providecommand{\eprint}[2][]{\url{#2}}

\bibitem[{\citenamefont{Senthil et~al.}(2004)\citenamefont{Senthil, Vishwanath,
  Balents, Sachdev, and Fisher}}]{senthil2004:science}
\bibinfo{author}{\bibfnamefont{T.}~\bibnamefont{Senthil}},
  \bibinfo{author}{\bibfnamefont{A.}~\bibnamefont{Vishwanath}},
  \bibinfo{author}{\bibfnamefont{L.}~\bibnamefont{Balents}},
  \bibinfo{author}{\bibfnamefont{S.}~\bibnamefont{Sachdev}}, \bibnamefont{and}
  \bibinfo{author}{\bibfnamefont{M.}~\bibnamefont{Fisher}},
  \bibinfo{journal}{Science} \textbf{\bibinfo{volume}{303}},
  \bibinfo{pages}{1490} (\bibinfo{year}{2004}).

\bibitem[{\citenamefont{Sandvik}(2007)}]{sandvik2007:deconf}
\bibinfo{author}{\bibfnamefont{A.~W.} \bibnamefont{Sandvik}},
  \bibinfo{journal}{Phys. Rev. Lett.} \textbf{\bibinfo{volume}{98}},
  \bibinfo{pages}{227202} (\bibinfo{year}{2007}).

\bibitem[{\citenamefont{Lou et~al.}(2009)\citenamefont{Lou, Sandvik, and
  Kawashima}}]{lou2009:sun}
\bibinfo{author}{\bibfnamefont{J.}~\bibnamefont{Lou}},
  \bibinfo{author}{\bibfnamefont{A.~W.} \bibnamefont{Sandvik}},
  \bibnamefont{and}
  \bibinfo{author}{\bibfnamefont{N.}~\bibnamefont{Kawashima}},
  \bibinfo{journal}{Phys. Rev. B} \textbf{\bibinfo{volume}{80}},
  \bibinfo{pages}{180414(R)} (\bibinfo{year}{2009}).

\bibitem[{\citenamefont{Kaul and Sandvik}(2012)}]{kaul2012:j1j2}
\bibinfo{author}{\bibfnamefont{R.~K.} \bibnamefont{Kaul}} \bibnamefont{and}
  \bibinfo{author}{\bibfnamefont{A.~W.} \bibnamefont{Sandvik}},
  \bibinfo{journal}{Phys. Rev. Lett.} \textbf{\bibinfo{volume}{108}},
  \bibinfo{pages}{137201} (\bibinfo{year}{2012}),
  \urlprefix\url{http://link.aps.org/doi/10.1103/PhysRevLett.108.137201}.

\bibitem[{\citenamefont{Motrunich and Vishwanath}(2004)}]{motrunich2004:hhog}
\bibinfo{author}{\bibfnamefont{O.~I.} \bibnamefont{Motrunich}}
  \bibnamefont{and}
  \bibinfo{author}{\bibfnamefont{A.}~\bibnamefont{Vishwanath}},
  \bibinfo{journal}{Phys. Rev. B} \textbf{\bibinfo{volume}{70}},
  \bibinfo{pages}{075104} (\bibinfo{year}{2004}).

\bibitem[{\citenamefont{Senthil et~al.}(2005)\citenamefont{Senthil, Balents,
  Sachdev, Vishwanath, and P.~A.~Fisher}}]{senthil:dccd}
\bibinfo{author}{\bibfnamefont{T.}~\bibnamefont{Senthil}},
  \bibinfo{author}{\bibfnamefont{L.}~\bibnamefont{Balents}},
  \bibinfo{author}{\bibfnamefont{S.}~\bibnamefont{Sachdev}},
  \bibinfo{author}{\bibfnamefont{A.}~\bibnamefont{Vishwanath}},
  \bibnamefont{and}
  \bibinfo{author}{\bibfnamefont{M.}~\bibnamefont{P.~A.~Fisher}},
  \bibinfo{journal}{Journal of the Physical Society of Japan}
  \textbf{\bibinfo{volume}{74}}, \bibinfo{pages}{1} (\bibinfo{year}{2005}),
  \eprint{https://doi.org/10.1143/JPSJS.74S.1},
  \urlprefix\url{https://doi.org/10.1143/JPSJS.74S.1}.

\bibitem[{\citenamefont{Gorshkov et~al.}(2010)\citenamefont{Gorshkov, Hermele,
  Gurarie, Xu, Julienne, Ye, Zoller, Demler, Lukin, and
  Rey}}]{gorshkov2010:sun}
\bibinfo{author}{\bibfnamefont{A.~V.} \bibnamefont{Gorshkov}},
  \bibinfo{author}{\bibfnamefont{M.}~\bibnamefont{Hermele}},
  \bibinfo{author}{\bibfnamefont{V.}~\bibnamefont{Gurarie}},
  \bibinfo{author}{\bibfnamefont{C.}~\bibnamefont{Xu}},
  \bibinfo{author}{\bibfnamefont{P.~S.} \bibnamefont{Julienne}},
  \bibinfo{author}{\bibfnamefont{J.}~\bibnamefont{Ye}},
  \bibinfo{author}{\bibfnamefont{P.}~\bibnamefont{Zoller}},
  \bibinfo{author}{\bibfnamefont{E.}~\bibnamefont{Demler}},
  \bibinfo{author}{\bibfnamefont{M.~D.} \bibnamefont{Lukin}}, \bibnamefont{and}
  \bibinfo{author}{\bibfnamefont{A.~M.} \bibnamefont{Rey}},
  \bibinfo{journal}{Nature Physics} \textbf{\bibinfo{volume}{6}},
  \bibinfo{pages}{289} (\bibinfo{year}{2010}), ISSN \bibinfo{issn}{1745-2473}.

\bibitem[{\citenamefont{Kaul}(2015)}]{kaulTriangular}
\bibinfo{author}{\bibfnamefont{R.~K.} \bibnamefont{Kaul}},
  \bibinfo{journal}{Phys. Rev. Lett.} \textbf{\bibinfo{volume}{115}},
  \bibinfo{pages}{157202} (\bibinfo{year}{2015}),
  \urlprefix\url{https://link.aps.org/doi/10.1103/PhysRevLett.115.157202}.

\bibitem[{\citenamefont{Block et~al.}(2020)\citenamefont{Block, D'Emidio, and
  Kaul}}]{kagome}
\bibinfo{author}{\bibfnamefont{M.~S.} \bibnamefont{Block}},
  \bibinfo{author}{\bibfnamefont{J.}~\bibnamefont{D'Emidio}}, \bibnamefont{and}
  \bibinfo{author}{\bibfnamefont{R.~K.} \bibnamefont{Kaul}},
  \bibinfo{journal}{Phys. Rev. B} \textbf{\bibinfo{volume}{101}},
  \bibinfo{pages}{020402(R)} (\bibinfo{year}{2020}),
  \urlprefix\url{https://link.aps.org/doi/10.1103/PhysRevB.101.020402}.

\bibitem[{\citenamefont{Demler et~al.}(2004)\citenamefont{Demler, Hanke, and
  Zhang}}]{demler:so5}
\bibinfo{author}{\bibfnamefont{E.}~\bibnamefont{Demler}},
  \bibinfo{author}{\bibfnamefont{W.}~\bibnamefont{Hanke}}, \bibnamefont{and}
  \bibinfo{author}{\bibfnamefont{S.-C.} \bibnamefont{Zhang}},
  \bibinfo{journal}{Rev. Mod. Phys.} \textbf{\bibinfo{volume}{76}},
  \bibinfo{pages}{909} (\bibinfo{year}{2004}),
  \urlprefix\url{https://link.aps.org/doi/10.1103/RevModPhys.76.909}.

\bibitem[{\citenamefont{Block and Kaul}(2012)}]{cubic}
\bibinfo{author}{\bibfnamefont{M.~S.} \bibnamefont{Block}} \bibnamefont{and}
  \bibinfo{author}{\bibfnamefont{R.~K.} \bibnamefont{Kaul}},
  \bibinfo{journal}{Phys. Rev. B} \textbf{\bibinfo{volume}{86}},
  \bibinfo{pages}{134408} (\bibinfo{year}{2012}),
  \urlprefix\url{https://link.aps.org/doi/10.1103/PhysRevB.86.134408}.

\bibitem[{\citenamefont{Sandvik}(2010)}]{sandvik2010:vietri}
\bibinfo{author}{\bibfnamefont{A.~W.} \bibnamefont{Sandvik}},
  \bibinfo{journal}{AIP Conference Proceedings}
  \textbf{\bibinfo{volume}{1297}}, \bibinfo{pages}{135} (\bibinfo{year}{2010}),
  \urlprefix\url{http://link.aip.org/link/?APC/1297/135/1}.

\bibitem[{\citenamefont{Swendsen and Wang}(1987)}]{swendsenwang}
\bibinfo{author}{\bibfnamefont{R.~H.} \bibnamefont{Swendsen}} \bibnamefont{and}
  \bibinfo{author}{\bibfnamefont{J.-S.} \bibnamefont{Wang}},
  \bibinfo{journal}{Phys. Rev. Lett.} \textbf{\bibinfo{volume}{58}},
  \bibinfo{pages}{86} (\bibinfo{year}{1987}),
  \urlprefix\url{https://link.aps.org/doi/10.1103/PhysRevLett.58.86}.

\bibitem[{\citenamefont{Wolff}(1989)}]{wolff:clusters}
\bibinfo{author}{\bibfnamefont{U.}~\bibnamefont{Wolff}},
  \bibinfo{journal}{Phys. Rev. Lett.} \textbf{\bibinfo{volume}{62}},
  \bibinfo{pages}{361} (\bibinfo{year}{1989}),
  \urlprefix\url{https://link.aps.org/doi/10.1103/PhysRevLett.62.361}.

\bibitem[{\citenamefont{Kirkpatrick et~al.}(1983)\citenamefont{Kirkpatrick,
  Gelatt, and Vecchi}}]{kirkpatrick:thermalannealing}
\bibinfo{author}{\bibfnamefont{S.}~\bibnamefont{Kirkpatrick}},
  \bibinfo{author}{\bibfnamefont{C.~D.} \bibnamefont{Gelatt}},
  \bibnamefont{and} \bibinfo{author}{\bibfnamefont{M.~P.}
  \bibnamefont{Vecchi}}, \bibinfo{journal}{Science}
  \textbf{\bibinfo{volume}{220}}, \bibinfo{pages}{671} (\bibinfo{year}{1983}),
  \eprint{https://www.science.org/doi/pdf/10.1126/science.220.4598.671},
  \urlprefix\url{https://www.science.org/doi/abs/10.1126/science.220.4598.671}.

\bibitem[{\citenamefont{Scherer}(2017)}]{scherer:computationalphysics}
\bibinfo{author}{\bibfnamefont{P.~O.~J.} \bibnamefont{Scherer}},
  \emph{\bibinfo{title}{Computational Physics, Simulation of Classical and
  Quantum Systems}} (\bibinfo{publisher}{Springer International Publishing AG},
  \bibinfo{address}{Gewerbestrasse 11, 6330 Cham, Switzerland},
  \bibinfo{year}{2017}), ISBN \bibinfo{isbn}{978-3-319-61087-0},
  \urlprefix\url{https://link.springer.com/book/10.1007/978-3-319-61088-7}.

\end{thebibliography}
\end{document}